\pdfoutput=1
\documentclass[11pt]{article}
\usepackage[margin=1in]{geometry}
\usepackage{graphicx}
\usepackage{cite}
\usepackage{makecell}
\usepackage{caption}
\usepackage{subcaption}

\def\Title#1{\begin{center} {\Large {\bf #1} } \end{center}}
\def\Author#1{\begin{center} {\normalsize {\sc #1} } \end{center}}
\def\Institution#1{\begin{center} {\normalsize {\it #1} } \end{center}}
\def\Abstract#1{\noindent {\normalsize {\bf Abstract:} {\normalfont #1}}}
\def\Conference{\vspace{4mm}\begin{raggedright} {\normalsize {\it Talk presented at the 2019 Meeting of the Division of Particles and Fields of the American Physical Society (DPF2019), July 29--August 2, 2019, Northeastern University, Boston, C1907293.} } \end{raggedright}\vspace{4mm}}

\begin{document}

%
%

\Title{Supervised learning of photoelectron counting in scintillator-based dark matter experiments}

\Author{Kolahal Bhattacharya, Christopher Jackson\\ For the MiniCLEAN Collaboration}

\Institution{Pacific Northwest National Laboratory, Richland WA-99354, USA}


\Abstract{Many scintillator based detectors employ a set of photomultiplier tubes (PMT) to observe the scintillation light from potential signal and background events. It is important to be able to count the number of photoelectrons (PE) in the pulses observed in the PMTs, because the position and energy reconstruction of the events is directly related to how well the spatial distribution of the PEs in the PMTs as well as their total number might be measured. This task is challenging for fast scintillators, since the PEs often overlap each other in time. Standard Bayesian statistics methods are often used and this has been the method employed in analyzing the data from liquid argon experiments such as MiniCLEAN and DEAP. In this work, we show that for the MiniCLEAN detector it is possible to use a multi-layer perceptron to learn the number of PEs using only raw pulse features with better accuracy and precision than existing methods. This can even help to perform position reconstruction with better accuracy and precision, at least in some generic cases.}

\Conference

%
%

\section{Introduction}
\begin{figure}[h]
  \centering
  \includegraphics[width=0.375\textwidth]{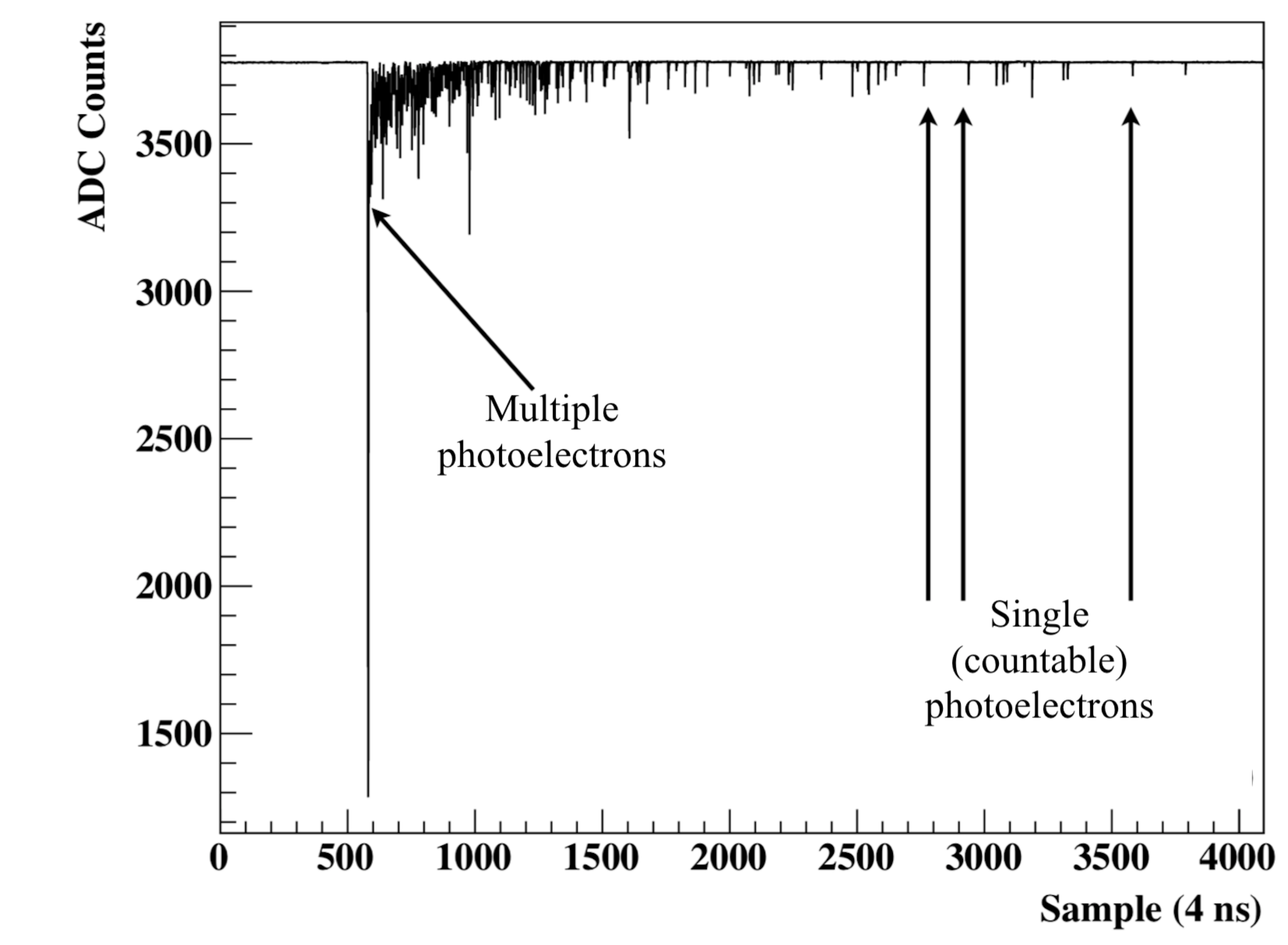}
  \caption{A typical pulse with photoelectron peaks overlapping during the prompt time scale of scintillation~\cite{pecountingplot}}
  \label{f1}
\end{figure}

Direct dark matter search experiments that are based on noble liquid targets read out by photomultiplier tubes (PMTs) often employ scintillation 
pulse shape discrimination methods to identify the particle causing the interaction. In this technique, the amount of light observed within 
the prompt time scale of scintillation (typically within $\sim100$ ns of the event 
trigger for argon) relative to the total amount of light observed in the event, is used 
to distinguish between nuclear and electronic recoil events. This quantity is termed as the prompt fraction, $f_p$, and is expressed as:
\begin{equation}
 f_p = \frac{\int_{t_0}^{t_{100}}V(t) dt}{\int_{t_0}^{t_f}V(t) dt},
\end{equation}
where $V(t)$ denotes the voltage waveform and the integral measures the light 
observed by the PMTs within the integral limits. The limits $t_0$, $t_{100}$ 
and $t_f$ denote the trigger time, first $100$ ns and total time window in the 
event for light collection. Figure~\ref{f1} shows that this integral is equivalent to counting the number of photoelectrons 
within the observed pulse. Clearly, the accuracy of $f_p$, which dictates the 
degree to which the nuclear recoil events and the electronic recoil events can 
be identified and separated, depends on how well the counting of photoelectrons 
can be performed. 

From Figure~\ref{f1}, it is clear that this is a difficult task 
at shorter time scales of scintillation where the photoelectrons tend to overlap 
and cannot be resolved with clarity. 
In the context of the MiniCLEAN experiment~\cite{akashi2019triplet}, a Bayesian statistics based counting method 
was developed~\cite{akashi2015improving}. This technique demonstrated better performance 
compared to an averaging technique that attempts to count the desired number 
by dividing the total charge deposited in a pulse by the mean charge per photoelectron. 
In this work, we intend to explore if we can further improve this limit to perform 
more accurate and precise counting of photoelectrons. This will benefit MiniCLEAN 
as well as other experiments in the global argon dark matter collaboration (e.g. DEAP-3600, 
DarkSide-20k etc.) which will employ pulse shape discrimination. It could potentially 
also improve the performance of any PMT-based single photon counting experiment. To this end, 
we carry out an exercise of supervised learning of the number of photoelectrons from 
the features of simulated PMT pulses.

\vspace{-0.25 cm}
\section{Description of supervised learning technique}
\begin{figure}[ht]
\centering
\begin{subfigure}{.5\textwidth}
  \centering
  \includegraphics[width=0.9\linewidth]{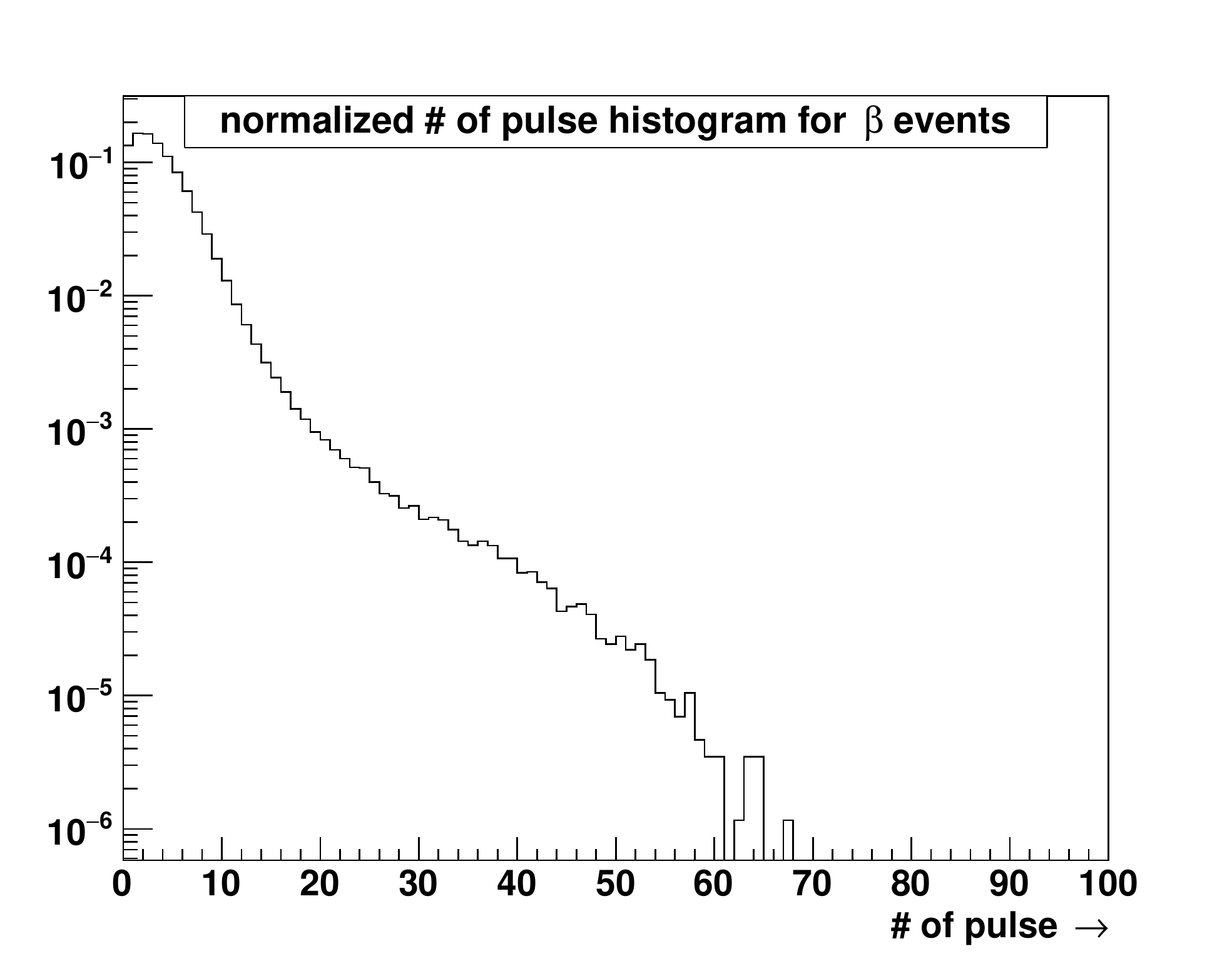}
  \caption{$\beta$}
  \label{ff_a}
\end{subfigure}%
\begin{subfigure}{.5\textwidth}
  \centering
  \includegraphics[width=.9\linewidth]{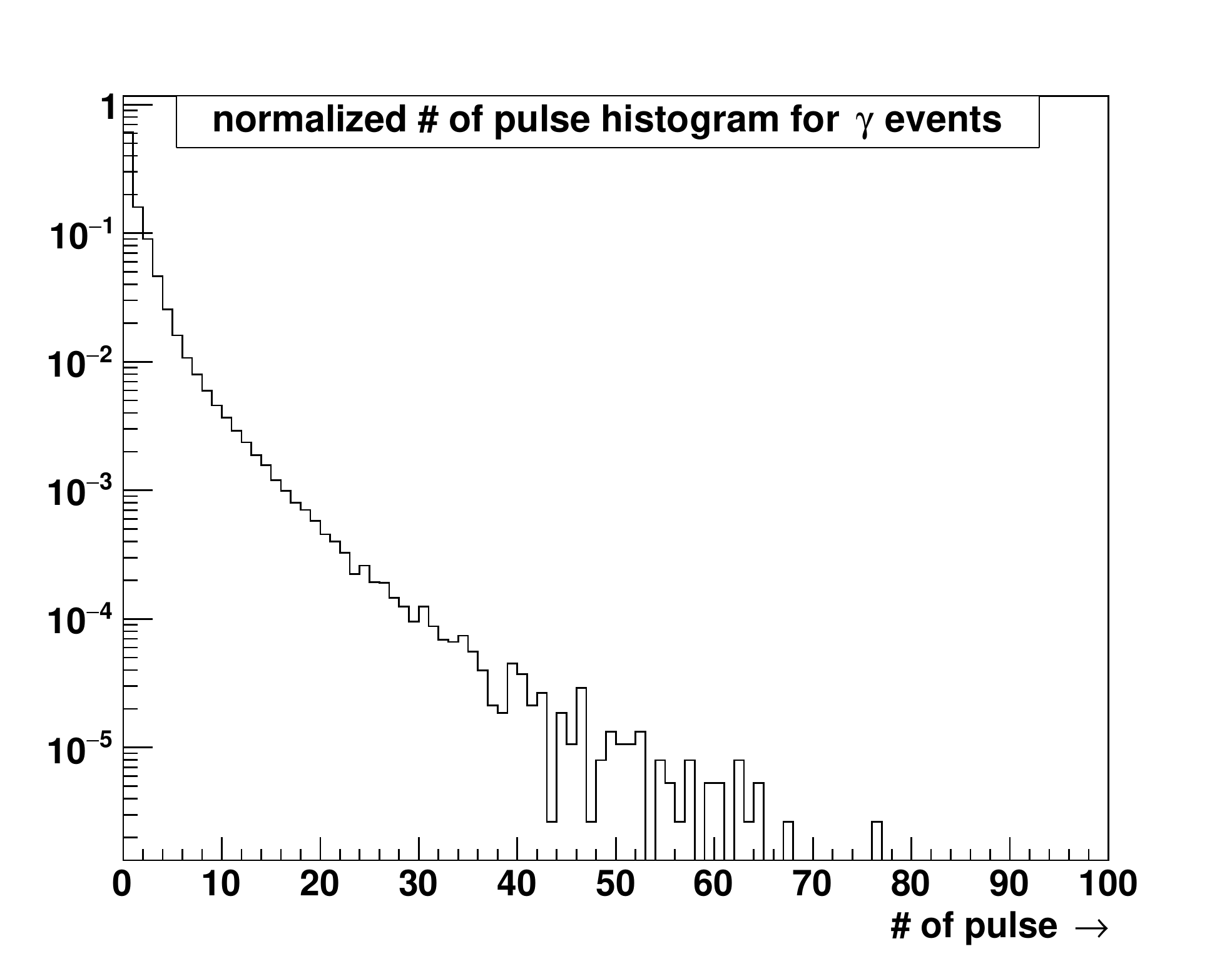}
  \caption{$\gamma$}
  \label{ff_b}
\end{subfigure}
\caption{Number of pulses distribution for (a) Argon-39 induced $\beta$ decay events and (b) $\gamma$ events}
\label{fig:ff}
\end{figure}

The data sample for this task has been generated using the RAT~\cite{rat-miniclean} 
simulation framework. The Monte Carlo program simulates the pulses generated by 
the PMTs and produces a realistic data sample in which all 92 PMTs of the MiniCLEAN 
detector, arranged centrally facing inside the spherical inner vessel, observe 
different number of pulses in an event. The waveforms are sampled into 4 nanosecond 
bins. We design a network that will learn the approximate number of photoelectrons 
observed by an individual PMT from the raw pulse features (mean time and width of the 
pulse waveform, the bin numbers containing the left and right edges of the pulse, the 
bin number where the peak of the pulse waveform is located, the charge contained within 
that bin and the total charge deposited in the pulse). Different events observe varying 
number of pulses, as seen in Figure~\ref{fig:ff}.

It is assumed that a PMT can observe at most 100 pulses in an event. The floating point numbers 
for all the pulses representing their features are written to a vector. We employ 
zero-padding when the true number of pulses is less than 100 in an event. Hence, 
a multi-layer perceptron with dense connections is trained on the number of Monte Carlo 
photoelectrons for all the PMTs in MiniCLEAN with about 3 
million events (training data) using the feature vector on an event by event basis. 
The detailed parameters of the network architecture are given in Table~\ref{t1} in the appendix~\ref{appendix}.

From the study of the PMTs in MiniCLEAN, it is observed that individual PMTs have 
different calibration coefficients and dark noise rates. Therefore, individual PMTs are trained 
in parallel in our work in the following manner. The feature vector prepared with the Monte Carlo 
sample for a given PMT is trained against the number of photoelectrons observed in that PMT 
itself, for all of the training data. The test data is a different Monte 
Carlo data sample with the same PMT calibration parameters. When the network prediction is seen to 
converge to the extent that the mean squared error between the network prediction and the true number 
of photoelectrons is very small ($<0.5$) and stable for an arbitrary unseen validation data set, the training 
is stopped and the predicted number from all the 92 PMTs are added up to give the predicted number of 
photoelectrons in an event. 

\section{Training characteristics}
\begin{figure}[htp]
\centering
\begin{subfigure}{.5\textwidth}
  \centering
  \includegraphics[width=0.95\linewidth]{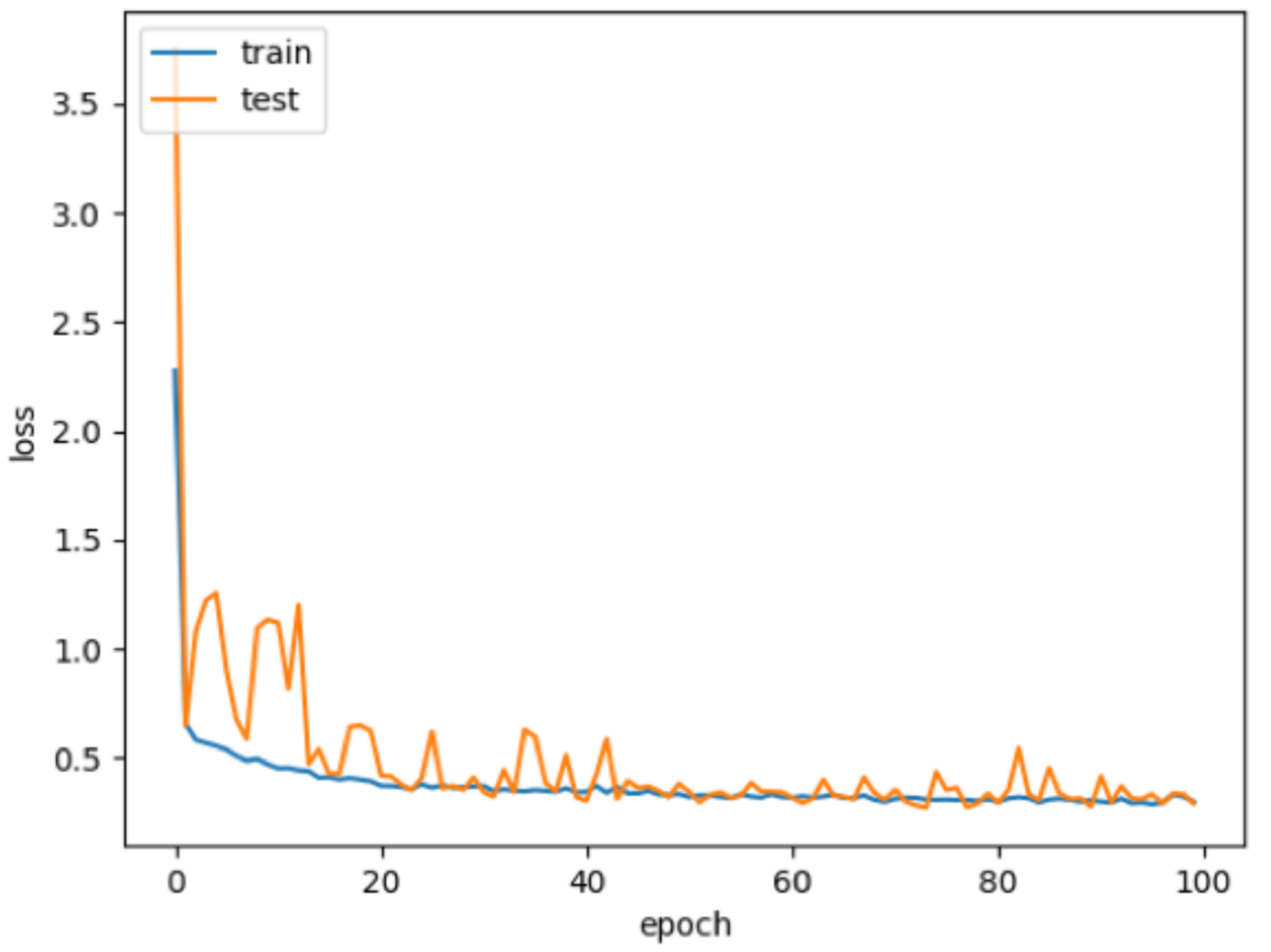}
  \caption{Loss vs. training epochs}
  \label{f2a}
\end{subfigure}%
\begin{subfigure}{.5\textwidth}
\vspace{0.35cm}
  \centering
  \includegraphics[width=.95\linewidth]{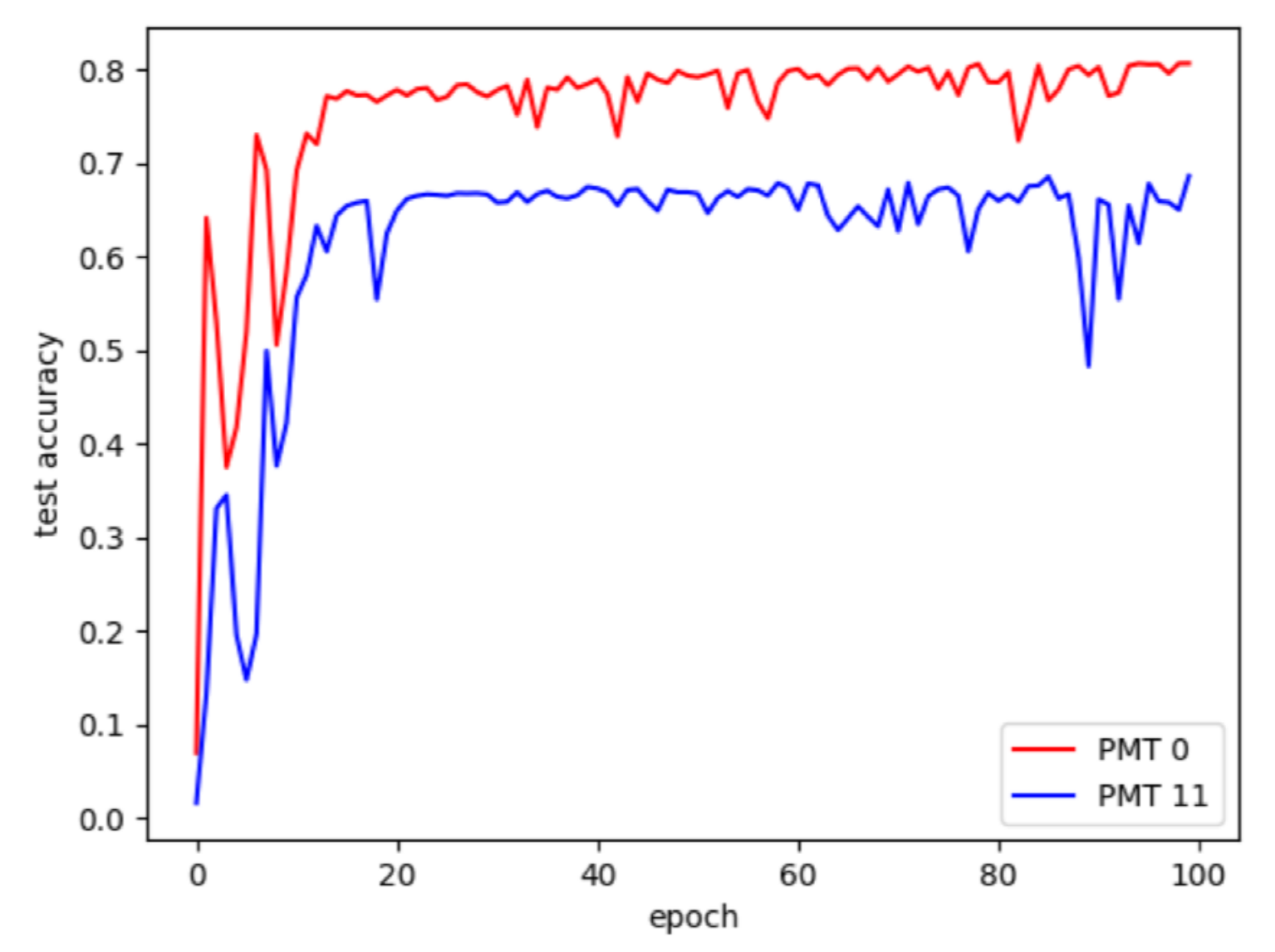}
  \caption{Accuracy for different PMTs vs. training epochs}
  \label{f2b}
\end{subfigure}
\caption{(a) The evolution of loss function observed during training the network; (b) different degrees 
   of accuracy in estimation achieved by PMT\#0 and PMT\#11. Similar characteristics are observed in training other PMTs as well.}
\label{fig:2}
\end{figure}

For a training session with Argon-39 $\beta$ decay Monte Carlo events, distributed uniformly throughout the 
detector, it is seen that the loss function becomes progressively smaller during training, as seen 
in Figure~\ref{f2a}. Also, there is no evidence of overtraining. It is observed that different PMTs 
exhibit different degrees of accuracy in estimating the correct number of photoelectrons (see Figure~\ref{f2b}), 
which is expected because each PMT has unique single photoelectron calibration parameters and dark 
hit rate.

\vspace{-0.5 cm}
\section{Outcome of network prediction}
\begin{figure}[htp]
\centering
\begin{subfigure}{.5\textwidth}
  \centering
  \includegraphics[width=0.95\linewidth]{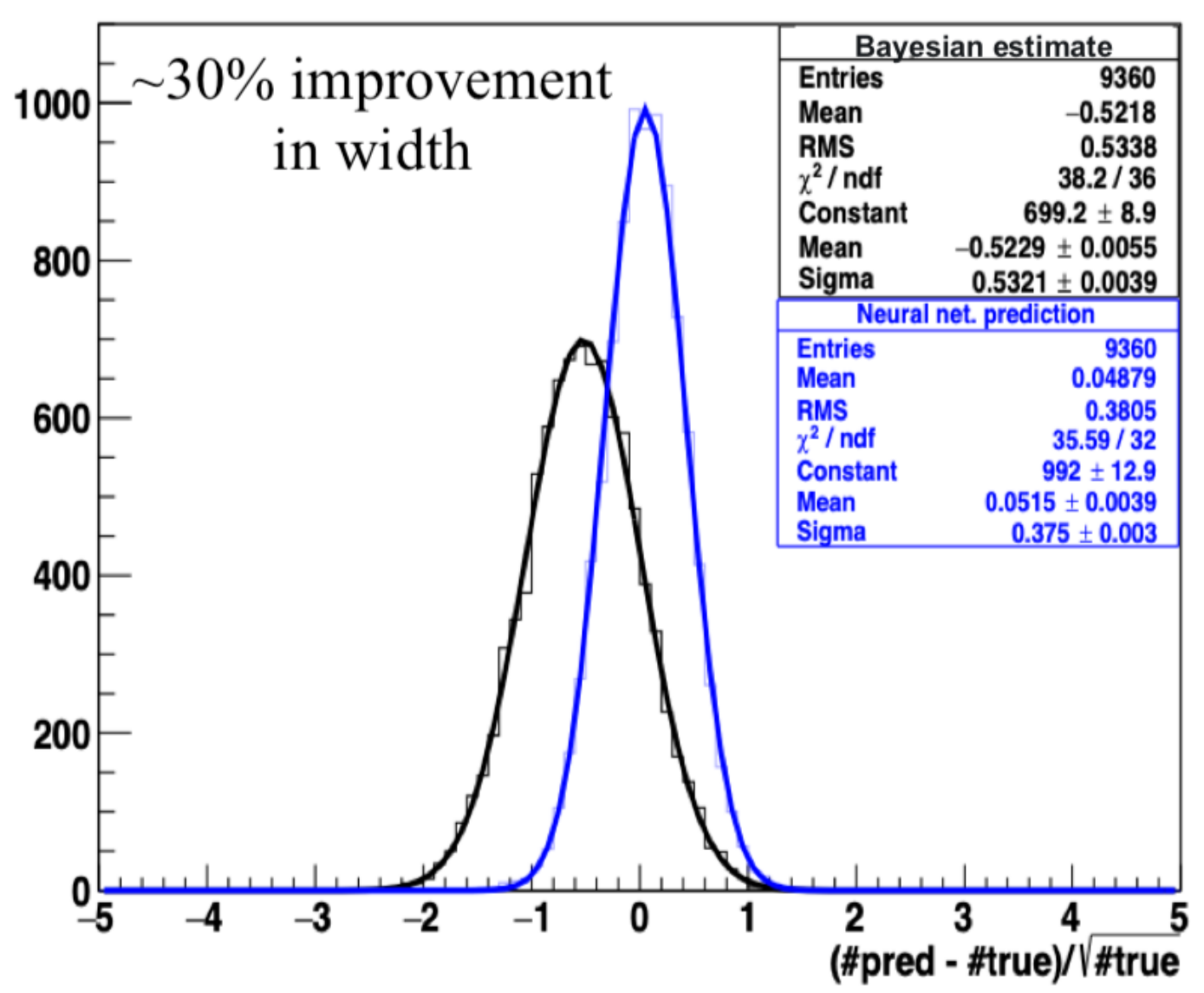}
  \caption{Comparison for $\beta$}
  \label{f3_a}
\end{subfigure}%
\begin{subfigure}{.5\textwidth}
  \centering
  \includegraphics[width=.95\linewidth]{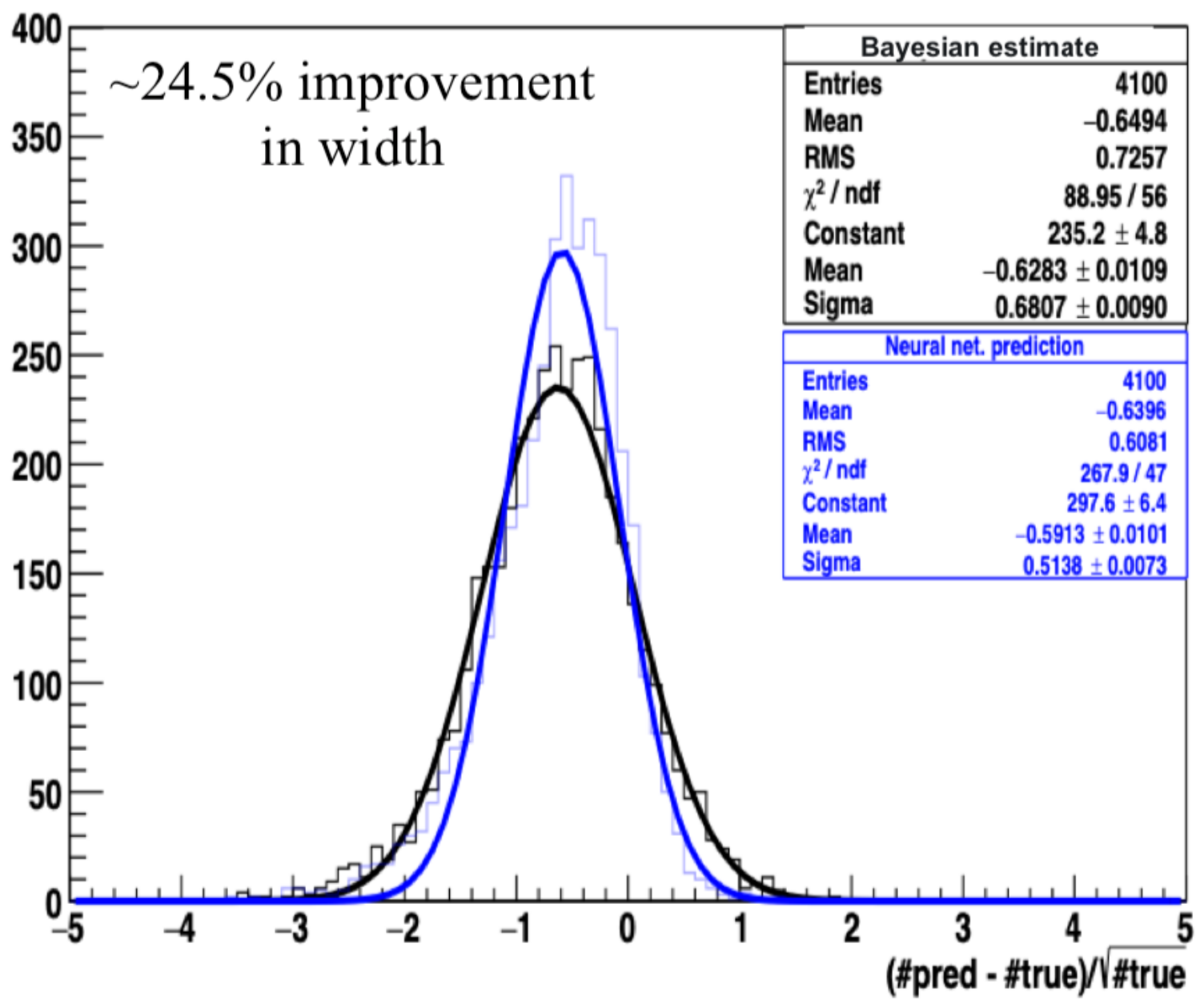}
  \caption{Comparison for $\gamma$}
  \label{f3_b}
\end{subfigure}
\caption{Comparison of performance of photoelectron counting between the network prediction and Bayeseian counting using pull distributions}
\label{fig:3}
\end{figure}

To evaluate if 
this approach of estimating the number of photoelectrons in an event is more accurate and precise 
compared to the Bayesian photoelectron counting~\cite{akashi2015improving}, we investigated the 
network prediction for the Monte Carlo data samples from Argon-39 $\beta$ 
and Uranium/Thorium chain $\gamma$ backgrounds (the latter mostly originate from the edges of the 
inner and outer vessels and the PMTs) and compared the pull distributions, 
defined as:
\begin{equation}
 \rm{pull}=\frac{estimated\ \#\ of\ photoelectrons-true\ \#\ of\ photoelectrons}{\sqrt{true\ \#\ of\ photoelectrons}},
\end{equation}
between the network prediction and Bayesian counting for identical sets of events. 
Figure~\ref{fig:3} shows the performance of the network, where the relative improvement 
in the precision of estimation (standard deviation of the pull distribution) with respect to 
the Bayesian counting method has been highlighted in percentage (e.g. $\sim30\%$ for 
Argon-39 $\beta$ events).

From Figure~\ref{fig:3}, it is clear that this method leads to a significant improvement 
in counting the number of photoelectrons in scintillator-PMT based experiments. 


\section{Application to vertex position reconstruction}
Since the network provides a means to measure the amount of light observed by different PMTs 
arranged at different angular positions around the inner vessel, it is possible to perform a 
position reconstruction of the event vertex by invoking a likelihood-based technique, using 
the distribution of observed photoelectrons across 
the detector. Substitution of this improved photoelectron counting algorithm led to 3.6\%, 12\% and 
14\% improvement in the precision of position estimation in $z$, $x$ and $y$ coordinates, respectively.



\section*{Acknowledgements}

The authors acknowledge the funding support from the NPAC initiative under the Laboratory Directed 
Research and Development (LDRD) program of Pacific Northwest National Laboratory. The research was 
performed using resources available through Research Computing at PNNL. The laboratory is operated 
by Battelle for the U.S. Department of Energy under Contract DE-AC05-76RL01830.

\appendix
\section{Network architecture}\label{appendix}

\begin{table}[h]
\begin{center}
\begin{tabular}{|c|c|c|c|c|} \hline 
{\bf{architecture}} & {\bf{weight initialization}} & {\bf{learning rate}} & {\bf{optimizer}} \\ \hline
 2 layers, 1024 neurons & Xavier initialization & \makecell{lr=0.5, $\beta_1=0.99$,\\ $\beta_2=0.999$, $\epsilon=1.e^{-8}$}  & Adam \\ \hline
{\bf{regularization}} & {\bf{loss function}} & {\bf{batch size}} & {\bf{activation function}} \\ \hline
batch normalization & mean squared error & 65,000 & ReLU \\ \hline
\end{tabular}
\caption{Parameters used for training a multi-layer perceptron to learn the number of photoelectrons observed by individual PMTs in MiniCLEAN}
\label{t1}
\end{center}
\end{table}

\bibliographystyle{unsrt}
\bibliography{DPF2019_template} 
\end{document}